\documentclass{article}

\usepackage{color}
\usepackage{amsmath, amssymb}

\newcommand{\ZM}{\mathbb{Z}}

\newcommand{\CM}{\mathbb{C}}
\newcommand{\PM}{\mathbb{P}}
\newtheorem{theorem}{Theorem}
\newtheorem{lemma}{Lemma} 
\newtheorem{prop}{Proposition} 
\newtheorem{cor}{Corollary}

\newcommand{\miniket}[1]{\vert#1\rangle}

\begin{document}

\title{{\bf IPS/Zeta Correspondence}
\vspace{15mm}}

\author{Takashi KOMATSU \\
Math. Research Institute Calc for Industry \\
Minami, Hiroshima, 732-0816, JAPAN \\ 
e-mail: ta.komatsu@sunmath-calc.co.jp
\\ \\ 
Norio KONNO \\
Department of Applied Mathematics, Faculty of Engineering \\ 
Yokohama National University \\
Hodogaya, Yokohama, 240-8501, JAPAN \\
e-mail: konno-norio-bt@ynu.ac.jp \\
\\ 
Iwao SATO \\ 
Oyama National College of Technology \\
Oyama, Tochigi, 323-0806, JAPAN \\ 
e-mail: isato@oyama-ct.ac.jp 
}

\date{\empty }

\maketitle

\vspace{50mm}


\vspace{20mm}










\clearpage

\begin{abstract}
Our previous works presented zeta functions by the Konno-Sato theorem or the Fourier analysis for one-particle models including random walks, correlated random walks, quantum walks, and open quantum random walks. This paper introduces a new zeta function for multi-particle models with probabilistic or quantum interactions, called the interacting particle system (IPS). We compute the zeta function for some tensor-type IPSs.
\end{abstract}

\vspace{10mm}

\begin{small}
\end{small}

\vspace{10mm}

\section{Introduction \label{sec01}}
In our previous paper \cite{KomatsuEtAl2021}, we studied a relation between the Grover walk and the zeta function based on the Konno-Sato theorem \cite{KonnoSato} and called this relation ``Grover/Zeta Correspondence". More precisely, we gave explicit formulas for the generalized zeta function and the generalized Ihara zeta function corresponding to the Grover walk with F-type and the positive-support version of the Grover walk with F-type on the vertex-transitive regular graph by the Konno-Sato theorem, respectively. The Grover walk is one of the most well-investigated quantum walks (QWs) inspired by the famous Grover algorithm. The QW is a quantum counterpart of the correlated random walk (CRW) which has the random walk (RW) as a special model. In fact, the CRW is the RW with memory. As for the QW, see \cite{Konno2008, ManouchehriWang, Portugal, Venegas} and as for the CRW and the RW, see \cite{Konno2009, Spitzer}, for example.

Furthermore, in the subsequent paper \cite{KomatsuEtAl2021b}, we extended the Grover walk with F-type and the positive-support version of the Grover walk with F-type to a class of walks with both F- and M-types by using not the Konno-Sato theorem but a method of the Fourier transform for the case of the $d$-dimensional torus. Our class contains QWs and CRWs. In addition, we treated the open quantum random walk (OQRW) which has the CRW as a special model. Concerning the OQRW, see \cite{AttalEtAl2012b, AttalEtAl2012a}, for example. We called this kind of the zeta function the {\em walk-type} zeta function and such a relationship ``Walk/Zeta Correspondence", corresponding to the above mentioned ``Grover/Zeta Correspondence". 

Our previous two papers \cite{KomatsuEtAl2021, KomatsuEtAl2021b} investigated one-particle models including RWs, CRWs, QWs, and OQRWs. This paper deals with  multi-particle models with probabilistic or quantum interactions, called the {\em interacting particle systems} (IPS) including probabilistic cellular automata (PCA) and quantum cellular automata (QCA). In particular, PCA have oriented percolation as a special model. We introduce the IPS-type zeta function and call ``IPS/Zeta Correspondence" corresponding to the above mentioned ``Grover/Zeta Correspondence" and ``Walk/Zeta Correspondence".

The rest of the present manuscript is organized as follows. In Section \ref{sec02}, we define the two-state discrete time IPS on the one-dimensional path space. Section \ref{sec03} gives some examples for PCA and QCA. In Section \ref{sec04}, we introduce the IPS-type zeta function. Moreover, we consider the zeta function for tensor model and generalized one in Sections \ref{sec05} and \ref{sec06} respectively. We treat especially two typical models defined by $Q^{(l)}_{QCA,1}(\xi_1,\xi_2)$ and $Q^{(l)}_{QCA,2}(\xi_1,\xi_2)$ for $\xi_1, \xi_2 \in [0, 2\pi)$. So our results are useful for investigating dynamics of IPS and would be helpful to study on computation of quantum information. Finally, Section \ref{sec07} is devoted to summary.

\section{Model \label{sec02}}
This section gives the definition of our models, i.e., IPSs. To do so, we first introduce the following notation: $\mathbb{Z}$ is the set of integers, $\mathbb{Z}_{\ge}$ is the set of non-negative integers, $\mathbb{Z}_{>}$ is the set of positive integers, and $\mathbb{C}$ is the set of complex numbers. 


Let $\mathbb{P}_N = \{ 0,1, \ldots, N-1 \}$ be the path space with $N$ sites. Throughout this paper, we mainly assume that $N \ge 2$. There are two states ``0" or ``1" at each site. Let $\eta (x) \in \{0,1\}$ denote the state of the site $x \in \PM_N$, i.e., $x=0,1,\ldots, N-1$. The configuration space is $\{0,1\}^{\mathbb{P}_N}$ with $2^N$ elements. Intuitively, a configuration $\eta = (\eta (0), \eta (1), \ldots, \eta (N-1)) \in \{0,1\}^{\PM_N}$ is given an occupation interpretation as follows: $\eta (x)=1$ means that a particle exists at site $x \in \PM_N$, and $\eta (x)=0$ means that $x$ is vacant. In this paper, we put
\begin{align}
\miniket 0 =
\begin{bmatrix}
1 \\
0
\end{bmatrix} 
,
\quad 
\miniket 1 =
\begin{bmatrix}
0 \\
1
\end{bmatrix}
.
\label{zeroone} 
\end{align}
For example, when $N=3$, a configuration $(0,0,1) \in \{0,1\}^{\mathbb{P}_3}$ means that the state ``0" at site 0, the state ``0" at site 1, and the state ``1" at site 2. In other words, $(\eta (0), \eta (1), \eta (2)) = (0,0,1)$. We also write $(0,0,1)$ by $\miniket 0 \miniket 0 \miniket 1 = \miniket 0 \otimes \miniket 0 \otimes \miniket 1$, where $\otimes$ means the tensor product. By using Eq. \eqref{zeroone}, we have
\begin{align*}
\miniket 0 \miniket 0 \miniket 1 = 
\begin{bmatrix}
1 \\
0
\end{bmatrix} 
\otimes
\begin{bmatrix}
1 \\
0
\end{bmatrix} 
\otimes
\begin{bmatrix}
0 \\
1
\end{bmatrix} 
=
\begin{bmatrix}
0 \\
1 \\
0 \\
0 \\
0 \\
0 \\
0 \\
0 
\end{bmatrix}
\in \CM^{2^3}.
\end{align*}

To define our model, we introduce the {\em local} operator $Q^{(l)}$ and the {\em global} operator $Q^{(g)}_N$ in the following way. This definition is based on Katori et al. \cite{KatoriEtAl2004}.

We first define the $4 \times 4$ matrix $Q^{(l)}$ by
\begin{align*}
Q^{(l)}
=
\begin{bmatrix}
a^{00}_{00} & a^{01}_{00} & a^{10}_{00} & a^{11}_{00} \\ 
a^{00}_{01} & a^{01}_{01} & a^{10}_{01} & a^{11}_{01} \\ 
a^{00}_{10} & a^{01}_{10} & a^{10}_{10} & a^{11}_{10} \\ 
a^{00}_{11} & a^{01}_{11} & a^{10}_{11} & a^{11}_{11}  
\end{bmatrix} 
,
\end{align*}
where $a^{ij}_{kl} \in \mathbb{C}$ for $i,j,k,l \in \{0,1\}.$  Let $\eta_n (x) \in \{0,1\}$ denote the state of the site $x \in \PM_N$ at time $n \in \ZM_{\ge}$. The element of $Q^{(l)}$, $a^{ij}_{kl}$, means the {\em transition weight} from $(\eta_n (x), \eta_n (x+1)) =(i,j)$ to $(\eta_{n+1} (x), \eta_{n+1} (x+1))=(k,l)$ for any $x=0,1, \ldots, N-2$ and $n \in \ZM_{\ge}$. If $a^{ij}_{kl} \in [0,1]$, then the transition weight can be the transition probability. We call ``$x$" the {\em left site} and ``$x+1$" the {\em right site}. Throughout this paper, we assume that $a^{ij}_{kl}=0$ if $j \not=l$. In other words, after the time transition, the state of the right site does not change. Therefore, under this assumption, $Q^{(l)}$ is rewritten as
\begin{align*}
Q^{(l)}
=
\begin{bmatrix}
a^{00}_{00} & \cdot & a^{10}_{00} & \cdot \\ 
\cdot & a^{01}_{01} & \cdot & a^{11}_{01} \\ 
a^{00}_{10} & \cdot & a^{10}_{10} & \cdot \\ 
\cdot & a^{01}_{11} & \cdot & a^{11}_{11}  
\end{bmatrix} 
,
\end{align*}
where ``$\cdot$" means 0. By definition, the interaction of our IPS is nearest neighbor. In particular, if $a^{ij}_{kl} \in \{0,1\}$, then the IPS is called the {\em cellular automaton} (CA). Next we define the $2^N \times 2^N$ matrix $Q^{(g)}_N$ by
\begin{align*}
Q^{(g)}_N
&= \left( I_2 \otimes I_2 \otimes \cdots \otimes I_2 \otimes Q^{(l)} \right) \left( I_2 \otimes I_2 \otimes \cdots \otimes Q^{(l)} \otimes I_2 \right) 
\\
& \qquad \cdots \left( I_2 \otimes Q^{(l)} \otimes \cdots \otimes I_2 \otimes I_2 \right) \left( Q^{(l)} \otimes I_2 \otimes \cdots \otimes I_2 \otimes I_2 \right),
\end{align*}
where $I_n$ is the $n \times n$ identity matrix. For example, if $N=3$, then the $2^3 \times 2^3$ matrix $Q^{(g)}_3$ is 
\begin{align*}
Q^{(g)}_3 
= \left( I_2 \otimes Q^{(l)} \right) \left( Q^{(l)} \otimes I_2 \right).
\end{align*}
If $N=4$, then the $2^4 \times 2^4$ matrix $Q^{(g)}_4$ is 
\begin{align*}
Q^{(g)}_4 
= \left( I_2 \otimes I_2 \otimes Q^{(l)} \right) \left( I_2 \otimes Q^{(l)} \otimes I_2 \right) \left( Q^{(l)} \otimes I_2 \otimes I_2 \right).
\end{align*}
Note that if $N=2$, then $Q^{(g)}_2 = Q^{(l)}$. Moreover, when $N=1$, we put $Q^{(g)}_1 = I_2$.

We see that when $N=4$, a transition weight from $(\eta_n (0), \eta_n (1), \eta_n (2), \eta_n (3)) =(i_0,i_1,i_2,i_3) \in \{0,1\}^4$ to $(\eta_{n+1} (0), \eta_{n+1} (1), \eta_{n+1} (2), \eta_{n+1} (3)) =(k_0,k_1,k_2,k_3) \in \{0,1\}^4$ is $a^{i_0 i_1}_{k_0 k_1} a^{i_1 i_2}_{k_1 k_2} a^{i_2 i_3}_{k_2 k_3}$ for any $n \in \ZM_{\ge}$, for instance.

The above mentioned model is called the {\em interacting particle systems} (IPS) in this paper. We consider two typical classes, one is {\em probabilistic cellular automata} (PCA) and the other is {\em quantum cellular automata} (QCA). Note that PCA is also called stochastic CA.

A model in PCA satisfies 
\begin{align}
a^{00}_{00} + a^{00}_{10} = a^{01}_{01} + a^{01}_{11} = a^{10}_{00} + a^{10}_{10} = a^{11}_{01} + a^{11}_{11} = 1, \quad a^{ij}_{kj} \in [0,1].
\label{condPCA}
\end{align}
That is, $Q^{(l)}$ becomes a transposed {\em stochastic matrix} (also called {\em transition matrix}). Furthermore, we easily see that `` $Q^{(l)}$ is a transposed stochastic matrix if and only if $Q^{(g)}_N$ is a transposed stochastic matrix". In other words, the sum of the elements of any column for $Q^{(l)}$ or $Q^{(g)}_N$ is equal to 1.

On the other hand, a model in QCA satisfies that $Q^{(l)}$ is unitary, i.e., 
\begin{align*}
&
|a^{00}_{00}|^2 + |a^{00}_{10}|^2 = |a^{01}_{01}|^2 + |a^{01}_{11}|^2 = |a^{10}_{00}|^2 + |a^{10}_{10}|^2 = |a^{11}_{01}|^2 + |a^{11}_{11}|^2 = 1,
\nonumber
\\
& 
\qquad a^{00}_{00} \  \overline{a^{10}_{00}} + a^{00}_{10} \  \overline{a^{10}_{10}} =
 a^{01}_{01} \  \overline{a^{11}_{01}} + a^{01}_{11} \  \overline{a^{11}_{11}} =0. 
\end{align*}
This QCA was introduced by Konno \cite{Konno2008b} as a quantum counterpart of the Domany-Kinzel model (DK model) \cite{DomanyKinzel1984}. The DK model is defined in Section \ref{sec03}. As in the case of the PCA, we easily see that `` $Q^{(l)}$ is unitary if and only if $Q^{(g)}_N$ is unitary".

The evolution of IPS on $\PM_N$ is determined by 
\begin{align*}
\eta_{n} = \left( Q^{(g)} _N \right)^n \eta_{0} \quad (n \in \ZM_{\ge})
\end{align*}
for an initial state $\eta_{0}$. Note that $\eta_{n}, \eta_{0} \in \CM^{2^N}$ and $Q^{(g)} _N$ is a $2^N \times 2^N$ matrix.

For example, when $N=3$ and $(\eta (0), \eta (1), \eta (2)) = (0,0,1) \in \CM^{2^3}$, we observe
\begin{align*}
Q^{(g)}_3 (0,0,1)
= a^{00}_{00} a^{01}_{01} (0,0,1) + a^{00}_{00} a^{01}_{11} (0,1,1) 
+ a^{00}_{10} a^{01}_{01} (1,0,1) + a^{00}_{10} a^{01}_{11} (1,1,1).  
\end{align*}
In the case of PCA, the probability that a configuration $(0,1,1)$ exists is $a^{00}_{00} a^{01}_{11}$. In the case of QCA, the probability that a configuration $(0,1,1)$ exists is $|a^{00}_{00} a^{01}_{11}|^2$.

\section{Example \label{sec03}}
In this section, we give some typical models.  The DK model in PCA is defined by
\begin{align*}
Q^{(l)}_{DK}
=
\begin{bmatrix}
1 & \cdot & 1-p & \cdot \\ 
\cdot & 1-p & \cdot & 1-q \\ 
\cdot & \cdot & p & \cdot \\ 
\cdot & p & \cdot & q  
\end{bmatrix} 
,
\end{align*}
where $p, \ q \in [0,1]$. In particular, if $q=p$ (resp. $q=1-(1-p)^2$), then the DK model becomes oriented site (resp. bond) percolation (see Durrett \cite{Durett1988}, Katori et al. \cite{KatoriEtAl2004}, for example). A generalized DK model is defined by
\begin{align*}
Q^{(l)}_{DK} (\xi_1, \xi_2, \xi_3, \xi_4)
=
\begin{bmatrix}
\cos ^2 \xi_1 & \cdot & \sin ^2 \xi_3 & \cdot \\ 
\cdot & \sin ^2 \xi_2 & \cdot & \cos ^2 \xi_4 \\ 
\sin ^2 \xi_1 & \cdot & \cos ^2 \xi_3 & \cdot \\ 
\cdot & \cos ^2 \xi_2 & \cdot & \sin ^2 \xi_4  
\end{bmatrix} 
,
\end{align*}
where $\xi_1, \xi_2, \xi_3, \xi_4 \in [0, 2 \pi)$. Then the original DK model is equivalent to
\begin{align*}
(\xi_1, \xi_2, \xi_3, \xi_4) = \left( 0, \xi_2, \xi_2, \xi_4 \right).
\end{align*}
The well-known Wolfram {\em Rule 90} (see \cite{Durett1988}) is 
\begin{align*}
(\xi_1, \xi_2, \xi_3, \xi_4) = \left( 0, 0, 0, 0 \right),
\end{align*}
that is,
\begin{align*}
Q^{(l)}_{DK} (0, 0, 0, 0)
=
\begin{bmatrix}
1 & \cdot & \cdot & \cdot \\ 
\cdot & \cdot & \cdot & 1 \\ 
\cdot & \cdot & 1 & \cdot \\ 
\cdot & 1 & \cdot & \cdot  
\end{bmatrix} 
.
\end{align*}
A trivial model is 
\begin{align*}
Q^{(l)}_{DK} (0, \pi/2, 0, \pi/2) = I_4.
\end{align*}

As for the model in QCA, we introduce two classes. One is given by
\begin{align*}
Q^{(l)}_{QCA,1} (\xi_1, \xi_2)
=
\begin{bmatrix}
\cos \xi_1 & \cdot & - \sin \xi_1 & \cdot \\ 
\cdot & \cos \xi_2 & \cdot & - \sin \xi_2 \\ 
\sin \xi_1 & \cdot & \cos \xi_1 & \cdot \\ 
\cdot & \sin \xi_2 & \cdot & \cos \xi_2  
\end{bmatrix} 
,
\end{align*}
where $\xi_1, \xi_2 \in [0, 2 \pi)$. In particular, $Q^{(l)}_{QCA,1} (0,0) = I_4$. The other is 
\begin{align*}
Q^{(l)}_{QCA,2} (\xi_1, \xi_2)
=
\begin{bmatrix}
\cos \xi_1 & \cdot & - \sin \xi_1 & \cdot \\ 
\cdot & - \sin \xi_2 & \cdot & \cos \xi_2 \\ 
\sin \xi_1 & \cdot & \cos \xi_1 & \cdot \\ 
\cdot & \cos \xi_2 & \cdot & \sin \xi_2  
\end{bmatrix} 
,
\end{align*}
where $\xi_1, \xi_2 \in [0, 2 \pi)$. In particular, $Q^{(l)}_{QCA,2} (0,0)$ becomes Rule 90. Therefore, Rule 90 is not only PCA but also QCA.

\section{Zeta function \label{sec04}}
Following the walk-type zeta function introduced by our previous paper \cite{KomatsuEtAl2021b}, we define the {\em IPS-type zeta function} by 
\begin{align}
\overline{\zeta} \left(Q^{(l)}, \PM_N, u \right) = \det \Big( I_{2^N} - u Q^{(g)}_N \Big)^{-1/2^N} \quad (N \in \ZM_{>}).
\label{satosan01ips}
\end{align}
Note that when $N=1$, we put $Q^{(g)}_1 = I_2$. Thus we have 
\begin{align}
\overline{\zeta} \left(Q^{(l)}, \PM_1, u \right) = (1-u)^{-1}.
\label{zetaN1}
\end{align}
The difference between the walk-type and IPS-type zeta functions is that the former is based on the one-particle on lattice space and the latter is based on multi-particle on configuration space. Concerning the zeta function, see \cite{KomatsuEtAl2021, KonnoSato}, for example.

Furthermore, we define $C_r (Q^{(l)}, \PM_N)$ by
\begin{align}
\overline{\zeta} \left(Q^{(l)}, \PM_N, u \right) = \exp \left( \sum_{r=1}^{\infty} \frac{C_r (Q^{(l)}, \PM_N)}{r} u^r \right) \quad (N \in \ZM_{>}).
\label{satosan03ips}
\end{align}
Note that when $N=1$, we see that Eq. \eqref{zetaN1} implies
\begin{align}
C_r (Q^{(l)}, \PM_1) = 1 \quad (r \in \ZM_{>}).
\label{tuika01}
\end{align}


Let $\lambda_j \ (j=1,2, \ldots, 2^N)$ be eigenvalues of $Q^{(g)}_N$. From Eq. \eqref{satosan01ips}, we have
\begin{align*}
\log \left\{ \overline{\zeta} \left(Q^{(l)}, \PM_N, u \right) \right\}
&=
- \frac{1}{2^N} \log \left\{ \det \Big( I_{2^N} - u Q^{(g)}_N \Big) \right\}
\\
&= 
- \frac{1}{2^N} \log \left\{ \prod_{j=1}^{2^N} \Big( 1 - u \lambda_j \Big) \right\}
\\
&= 
- \frac{1}{2^N} \sum_{j=1}^{2^N} \log \Big( 1 - u \lambda_j \Big)
\\
&= 
\sum_{r=1}^{\infty} \left( \frac{1}{2^N} \sum_{j=1}^{2^N} \lambda_j^r \right) \frac{u^r}{r}
\\
&= 
\sum_{r=1}^{\infty} \left\{ \frac{1}{2^N} {\rm tr} \left( \left( Q^{(g)}_N \right)^r \right) \right\} \frac{u^r}{r},
\end{align*}
where ${\rm tr} (A)$ denotes the trace of a square matrix $A$. Thus we get
\begin{align}
\log \left\{ \overline{\zeta} \left(Q^{(l)}, \PM_N, u \right) \right\} = \sum_{r=1}^{\infty} \left\{ \frac{1}{2^N} {\rm tr} \left( \left( Q^{(g)}_N \right)^r \right) \right\} \frac{u^r}{r}.
\label{satosan03bips}
\end{align}

Combining Eq. \eqref{satosan03ips} with Eq. \eqref{satosan03bips} implies the following general result.
\begin{prop}
\begin{align*}
C_r (Q^{(l)}, \PM_N) = \frac{1}{2^N} {\rm tr} \left( \left( Q^{(g)}_N \right)^r \right) \quad (r, N \in \ZM_{>}).
\end{align*}
\label{propips}
\end{prop}

\section{Tensor model \label{sec05}}
In this section, we introduce a model whose $Q^{(l)}$ is defined by
\begin{align}
Q^{(l)} = Q^{(l,1)} \otimes Q^{(l,2)},
\label{tensor01}
\end{align}
where $Q^{(l,1)}$ and $Q^{(l,2)}$ are $2 \times 2$ matrices. This model is called the {\em tensor model} here. To compute $C_r$ for a general $N$ case, we begin with $r=1$ and $N=3$. Then, from Eq. \eqref{tensor01}, we see 
\begin{align*}
Q^{(g)}_3 
&= \left( I_2 \otimes Q^{(l)} \right) \left( Q^{(l)} \otimes I_2 \right)
\\
&= \left( I_2 \otimes Q^{(l,1)} \otimes Q^{(l,2)} \right) \left( Q^{(l,1)} \otimes Q^{(l,2)} \otimes I_2 \right)
\\
&= Q^{(l,1)} \otimes Q^{(l,1)} Q^{(l,2)} \otimes Q^{(l,2)}.
\end{align*}
By using ${\rm tr} (A \otimes B) = {\rm tr} (A) {\rm tr} (B)$, we have 
\begin{align*}
{\rm tr} \left( Q^{(g)}_3 \right) = {\rm tr} \left( Q^{(l,1)} \right) {\rm tr} \left( Q^{(l,1)} Q^{(l,2)} \right) {\rm tr} \left( Q^{(l,2)} \right).
\end{align*}
Similarly, for a general $r$, we get 
\begin{align*}
{\rm tr} \left( \left( Q^{(g)}_3 \right)^r \right) = {\rm tr} \left( \left( Q^{(l,1)}  \right)^r \right) {\rm tr} \left( \left( Q^{(l,1)} Q^{(l,2)} \right)^r \right) {\rm tr} \left( \left( Q^{(l,2)} \right)^r \right).
\end{align*}
Moreover, for a general $N$, we see
\begin{align*}
{\rm tr} \left( \left( Q^{(g)}_N \right)^r \right) = {\rm tr} \left( \left( Q^{(l,1)}  \right)^r \right) \left\{ {\rm tr} \left( \left( Q^{(l,1)} Q^{(l,2)} \right)^r \right) \right\}^{N-2} {\rm tr} \left( \left( Q^{(l,2)} \right)^r \right).
\end{align*}
By Proposition \ref{propips}, we have the following key result for the tensor model.
\begin{prop}
For any $r=1,2, \ldots$ and $N=2,3, \ldots$, we obtain
\begin{align*}
C_r (Q^{(l)}, \PM_N) = \frac{1}{2^N} {\rm tr} \left( \left( Q^{(l,1)}  \right)^r \right) \left\{ {\rm tr} \left( \left( Q^{(l,1)} Q^{(l,2)} \right)^r \right) \right\}^{N-2} {\rm tr} \left( \left( Q^{(l,2)} \right)^r \right).
\end{align*}
\label{proptm}
\end{prop}
Next we will find out a necessary and sufficient condition for Eq. \eqref{tensor01}. To do so, we put
\begin{align*}
Q^{(l,1)}
=
\begin{bmatrix}
a & b \\
c & d 
\end{bmatrix}
,
\quad
Q^{(l,2)}
=
\begin{bmatrix}
e & f \\
g & h 
\end{bmatrix}
,
\end{align*}
where $a, \ b, \ldots, \ h \in \CM.$ Furthermore, we assume that 
\begin{align}
Q^{(l,1)} \not= O_2, \qquad Q^{(l,2)} \not= O_2, 
\label{tensor03}
\end{align}
where $O_n$ is the $n \times n$ zero matrix. Then a direct computation implies
\begin{lemma}
Under the condition \eqref{tensor03}, $(f,g) = (0,0)$ is a necessary and sufficient condition for
\begin{align*}
Q^{(l)} = Q^{(l,1)} \otimes Q^{(l,2)}.
\end{align*}
\label{tensorlemma}
\end{lemma}
In fact, if $(f,g) = (0,0)$, then we have
\begin{align*}
Q^{(l)} 
= Q^{(l,1)} \otimes Q^{(l,2)} 
=
\begin{bmatrix}
a & b \\
c & d 
\end{bmatrix}
\otimes
\begin{bmatrix}
e & 0 \\
0 & h 
\end{bmatrix}
=
\begin{bmatrix}
ae & \cdot & be & \cdot \\
\cdot & ah & \cdot & bh \\
ce & \cdot & de & \cdot \\
\cdot & ch & \cdot & dh 
\end{bmatrix}
.
\end{align*}
In other words, if the state of the right site is ``0", then we multiply a weight ``$e$" and if the state of the right site is ``1", then we multiply a weight ``$h$", such as ``$(0,0) \to (1,0)$" with transition weight ``$ce$" and ``$(0,1) \to (1,1)$" with transition weight ``$ch$".  

Combining Proposition \ref{propips} with Proposition \ref{proptm} gives one of our main results.

\begin{theorem}
We assume that $N \ge 2$. We consider the tensor model with 
\begin{align*}
Q^{(l)} 
= Q^{(l,1)} \otimes Q^{(l,2)} 
=
\begin{bmatrix}
a & b \\
c & d 
\end{bmatrix}
\otimes
\begin{bmatrix}
e & 0 \\
0 & h 
\end{bmatrix}
.
\end{align*}
Let $\lambda_{+}, \ \lambda_{-}$ denote the eigenvalues of $Q^{(l,1)}$ and $\widetilde{\lambda}_{+}, \ \widetilde{\lambda}_{-}$ denote the eigenvalues of $Q^{(l,1)}  Q^{(l,2)}$. Then we obtain
\begin{align*}
C_r (Q^{(l)}, \PM_N) = \frac{1}{2^N} \left( \lambda_{+}^r + \lambda_{-}^r \right) \left( \widetilde{\lambda}_{+}^r + \widetilde{\lambda}_{-}^r \right)^{N-2} \left( e^r + h^r \right) \quad (r \in \ZM_{>}).
\end{align*}
In particular, if $e=h=1$, i.e., $Q^{(l,2)}=I_2$, then we have
\begin{align}
C_r (Q^{(l)}, \PM_N) = \frac{1}{2^{N-1}} \left( \lambda_{+}^r + \lambda_{-}^r \right)^{N-1} \quad (r \in \ZM_{>}).
\label{eqtmtm2}
\end{align}
\label{tmtm}
\end{theorem}

We begin with a trivial model with $Q^{(l,1)} = Q^{(l,2)} = I_2$, i.e., $Q^{(l)} =I_4$. In this case, we see that $\lambda_{+} = \lambda_{-}=1$. Thus, from  Eq. \eqref{eqtmtm2}, we get 
\begin{align*}
C_r (I_4, \PM_N) = \frac{1}{2^{N-1}} \left( 1^r + 1^r \right)^{N-1} = \frac{1}{2^{N-1}} \times 2^{N-1} = 1.
\end{align*}
Remark that when $N=1$, we see that Eq. \eqref{tuika01} gives $C_r (I_4, \PM_1) = 1$ for $r \in \ZM_{>}$. So we have
\begin{cor}
\begin{align}
&
C_r (I_4, \PM_N) = 1 \quad (r, N \in \ZM_{>}),
\label{trivialmodel}
\\
&\log \left( \overline{\zeta} \left(I_4, \PM_N, u \right)^{-1} \right) = \log (1-u) \quad (N \in \ZM_{>}).  
\label{trivialmodelzeta}
\end{align}
\end{cor}
We should note that Eq. \eqref{trivialmodelzeta} implies that all eigenvalues of $Q^{(g)}_N$ are 1.

If we consider PCA, then Eq. \eqref{condPCA} gives 
\begin{align*}
(a+c)e=(a+c)h=(b+d)e=(b+d)h=1.
\end{align*}
So we have $e=h(\not=0)$. This means that the state of the right site does not affect the transition weight. In other words, this IPS does not have the interaction. 

Next we consider QCA. Then we can find out a non-trivial model such as\begin{align*}
Q^{(l)} 
&= Q^{(l,1)} \otimes Q^{(l,2)} 
=
\begin{bmatrix}
\cos \xi & - \sin \xi \\
\sin \xi & \cos \xi 
\end{bmatrix}
\otimes
\begin{bmatrix}
e^{i \theta_1} & 0 \\
0 & e^{i \theta_2}
\end{bmatrix}
\\
&=
\begin{bmatrix}
e^{i \theta_1} \cos \xi & \cdot & - e^{i \theta_1} \sin \xi & \cdot \\ 
\cdot & e^{i \theta_2} \cos \xi & \cdot & - e^{i \theta_2} \sin \xi \\ 
e^{i \theta_1} \sin \xi & \cdot & e^{i \theta_1} \cos \xi & \cdot \\ 
\cdot & e^{i \theta_2} \sin \xi & \cdot & e^{i \theta_2} \cos \xi  
\end{bmatrix} 
,
\end{align*}
where $\xi, \theta_1, \theta_2 \in [0, 2 \pi)$. When $(\theta_1, \theta_2) = (0,0)$, i.e., $Q^{(l,2)} =I_2$, this $Q^{(l)}$ becomes $Q^{(l)}_{QCA,1} (\xi, \xi)$.  In this case, we have
\begin{align*}
\lambda_{+} = e^{i \xi}, \qquad \lambda_{-} = e^{- i \xi}.
\end{align*}
From Eq. \eqref{eqtmtm2}, we get
\begin{align}
C_r (Q^{(l)}_{QCA,1} (\xi, \xi), \PM_N) 
= \frac{1}{2^{N-1}} \left( e^{i \xi r} +  e^{-i \xi r} \right)^{N-1} 
= \left( \cos \left( \xi r \right) \right)^{N-1}.
\label{tuika02}
\end{align}
Let $\{ T_n (x) \}$ denote the Chebychev polynomials of the first kind (see Andrews et al. \cite{Andrews1999}):
\begin{align*}
T_n (x) = \cos \left( n \cdot \cos^{-1} (x) \right) \qquad \left(n \in \ZM_{\ge}, \ x \in [-1,1] \right),
\end{align*}
where $\cos^{-1} (x) = {\rm arccos} (x)$. For example, 
\begin{align*}
T_0 (x) = 1, \quad T_1 (x) = x, \quad T_2 (x) = 2x^2-1, \quad T_3 (x) = 4x^3-3x, \ldots.
\end{align*}
Note that $T_n (1) = 1$ for any $n \in \ZM_{\ge}$ . Therefore, we obtain 
\begin{cor}
\begin{align}
C_r (Q^{(l)}_{QCA,1} (\xi, \xi), \PM_N) = T_r (\cos \xi)^{N-1}
\label{QCA2cor}
\end{align}
for $r, N \in \ZM_{>}$ and $\xi \in [0,2 \pi)$. 
\end{cor}
In particular, if we take $\xi=0$, then Eq. \eqref{QCA2cor} gives 
\begin{align}
C_r (Q^{(l)}_{QCA,1} (0, 0), \PM_N) = 1 \quad (r, N \in \ZM_{>}).
\label{QCA2cortri}
\end{align}
On the other hand, noting that $Q^{(l)}_{QCA,1} (0, 0) = I_4$, we confirm that Eq. \eqref{QCA2cortri} is consistent with Eq. \eqref{trivialmodel}.

We will compute the zeta function for $Q^{(l)}_{QCA,1} (\xi, \xi)$. From Eq. \eqref{tuika02}, we see
\begin{align*}
- \log \left\{ \overline{\zeta} \left(Q^{(l)}_{QCA,1}(\xi,\xi), \PM_N, u \right)^{-1} \right\} 
&
= \sum_{r=1}^{\infty} \frac{C_r (Q^{(l)}_{QCA,1}(\xi,\xi), \PM_N)}{r} u^r 
\\
&
= \sum_{r=1}^{\infty} \frac{1}{r} \left( \frac{e^{ir \xi} + e^{-ir \xi}}{2} \right)^{N-1} u^r 
\\
&
= \frac{1}{2^{N-1}} \sum_{r=1}^{\infty} \frac{1}{r} \sum_{k=0}^{N-1} {N-1 \choose k} (e^{ir \xi})^k (e^{-ir \xi})^{N-1-k} \ u^r 
\\
&
= \frac{1}{2^{N-1}} \sum_{k=0}^{N-1} {N-1 \choose k} \sum_{r=1}^{\infty} \frac{1}{r} \left\{ e^{i (2k- (N-1)) \xi} u \right\}^r
\\
&
= - \frac{1}{2^{N-1}} \sum_{k=0}^{N-1} {N-1 \choose k} \log \left\{ 1 - e^{i (2k- (N-1)) \xi} u \right\}.
\end{align*}
Therefore the following result is obtained.
\begin{theorem}
We assume that $N \in \ZM_{>}$. Then we have
\begin{align*}
\log \left\{ \overline{\zeta} \left(Q^{(l)}_{QCA,1}(\xi,\xi), \PM_N, u \right)^{-1} \right\} = \frac{1}{2^{N-1}} \sum_{k=0}^{N-1} {N-1 \choose k} \log \left\{ 1 - e^{i (2k- (N-1)) \xi} u \right\}
\end{align*}
for $\xi \in [0,2 \pi)$. 
\label{zeta}
\end{theorem}
When $N=1$, we see that Eq. \eqref{zetaN1} gives $\overline{\zeta} \left(Q^{(l)}_{QCA,1}(\xi,\xi), \PM_1, u \right)^{-1}= 1-u$. So we confirm that Theorem \ref{zeta} holds for $N=1$. Next we will consider $N \to \infty$ case. To do so, we introduce a symmetric simple random walk at time $n$ on $\ZM$, denoted by $S_n$, whose distribution is given by
\begin{align}
P \Big( S_n = 2k -n \Big) = \frac{1}{2^{n}} {n \choose k} 
\label{symrw}
\end{align}
for $k=0,1, \ldots, n$. Combining Theorem \ref{zeta} with Eq. \eqref{symrw} implies 
\begin{align*}
&\log \left\{ \overline{\zeta} \left(Q^{(l)}_{QCA,1}(\xi,\xi), \PM_N, u \right)^{-1} \right\} 
\\
& \qquad \qquad = \sum_{k=0}^{N-1} \log \Big\{ 1 - \exp \Big( i (2k- (N-1)) \xi \Big) \cdot u \Big\} \  P \Big( S_{N-1} = 2k - (N-1) \Big)
\\
& \qquad \qquad = E \Big[ \log \Big\{ 1 - \exp \big( i S_{N-1} \xi \big) \cdot u \Big\} \Big].
\end{align*}
By using this, we get
\begin{align*}
&\log \left\{ \overline{\zeta} \left(Q^{(l)}_{QCA,1}(\xi/\sqrt{N},\xi/\sqrt{N}), \PM_N, u \right)^{-1} \right\} 
\\
& \qquad \qquad = E \left[ \log \left\{ 1 - \exp \left( i \frac{S_{N-1}}{\sqrt{N-1}} \frac{\sqrt{N-1}}{\sqrt{N}} \ \xi \right) \cdot u \right\} \right]
\\
& \qquad \qquad \to \quad E \Big[ \log \Big( 1 - e^{i \xi Z} \cdot u \Big) \Big] \qquad (N \to \infty),
\end{align*}
where $Z$ is the normal distribution with mean $0$ and variance $1$. Here we used the Central Limit Theorem:
\begin{align*}
\frac{S_{n}}{\sqrt{n}} \quad \Rightarrow \quad Z \qquad (n \to \infty),
\end{align*}
where $\Rightarrow$ means weak convergence (see Spitzer \cite{Spitzer}). Thus we obtain
\begin{cor}
\begin{align*}
\lim_{N \to \infty} \log \left\{ \overline{\zeta} \left(Q^{(l)}_{QCA,1}(\xi/\sqrt{N},\xi/\sqrt{N}), \PM_N, u \right)^{-1} \right\} 
= E \Big[ \log \Big( 1 - e^{i \xi Z} \cdot u \Big) \Big]
\end{align*}
for $\xi \in [0,2 \pi)$. Here $Z$ is the normal distribution with mean $0$ and variance $1$. 
\label{zetacor}
\end{cor}

\section{Generalized tensor model \label{sec06}}
In the previous section, we considered a tensor model with $Q^{(l)} = Q^{(l,1)} \otimes Q^{(l,2)}$, in particular, 
\begin{align*}
Q^{(l)}_{QCA,1} (\xi, \xi)
=
\begin{bmatrix}
\cos \xi & - \sin \xi \\
\sin \xi & \cos \xi 
\end{bmatrix}
\otimes
I_2
\qquad (\xi \in [0, 2 \pi)).
\end{align*}
This section deals with a generalized tensor model with $Q^{(l)} = Q^{(l,1)}_1 \otimes Q^{(l,2)}_1 + Q^{(l,1)}_2 \otimes Q^{(l,2)}_2$. As a typical example, we will treat the following model:
\begin{align}
Q^{(l)}_{QCA,2} (0, \xi)
=
I_2 \otimes E_{00} + 
\sigma (\xi)
\otimes
E_{11}
\qquad (\xi \in [0, 2 \pi)),
\label{oto01}
\end{align}
where 
\begin{align*}
\sigma (\xi) = 
\begin{bmatrix}
- \sin \xi & \cos \xi \\
\cos \xi & \sin \xi 
\end{bmatrix}
, \quad
E_{00} =
\begin{bmatrix}
1 & 0 \\
0 & 0 
\end{bmatrix}
, \quad
E_{11} =
\begin{bmatrix}
0 & 0 \\
0 & 1 
\end{bmatrix}
.
\end{align*}
Note that
\begin{align*}
\sigma (\xi)^2 = I_2 \qquad (\xi \in [0, 2 \pi)).
\end{align*}
From now on, we put 
\begin{align*}
Q^{(l)} (\xi) = Q^{(l)}_{QCA,2} (0, \xi).
\end{align*}
Then we obtain
\begin{lemma}
\begin{align}
& 
Q^{(g)} _{N+1} (\xi) 
= Q^{(g)} _{N} (\xi) \otimes E_{00} + Q^{(g), \sigma} _{N} (\xi) \otimes E_{11},\label{lemoto01}
\\
&
\left( Q^{(g)} _{N+1} (\xi) \right)^r 
= \left( Q^{(g)} _{N} (\xi) \right)^r \otimes E_{00} + \left( Q^{(g), \sigma} _{N} (\xi) \right)^r \otimes E_{11},
\label{lemoto02}
\\
&
{\rm tr} \left( \left( Q^{(g)} _{N+1} (\xi) \right)^r \right) 
= {\rm tr} \left( \left( Q^{(g)} _{N} (\xi) \right)^r \right) + {\rm tr} \left( \left( Q^{(g), \sigma} _{N} (\xi) \right)^r \right)
\label{lemoto03}
\end{align}
for $N, r \in \ZM_{>}$ and $\xi \in [0, 2 \pi)$. Here
\begin{align*}
&
Q^{(g), \sigma} _{N} (\xi) = \sigma_N (\xi) Q^{(g)} _{N} (\xi), \qquad  \sigma_N  (\xi) = I_2 ^{\otimes (N-1)} \otimes \sigma (\xi),
\\
&
Q^{(g)} _{1} (\xi) = I_2, \qquad  Q^{(g), \sigma} _{1} (\xi) = \sigma (\xi),
\end{align*}
where $A^{\otimes n} = A \otimes \cdots \otimes A$ with $n$ terms.
\label{lemma02}
\end{lemma}
{\bf Proof}. We begin with 
\begin{align*}
Q^{(g)} _{N+1} (\xi) 
&= \left( I_2^{\otimes (N-1)} \otimes Q^{(l)} (\xi) \right) \left( I_2^{\otimes (N-2)} \otimes Q^{(l)} (\xi) \otimes I_2 \right) 
\\
& \qquad \cdots \left( I_2 \otimes Q^{(l)} (\xi) \otimes I_2^{\otimes (N-2)} \right) \left( Q^{(l)} (\xi) \otimes I_2^{\otimes (N-1)} \right)
\\
&= \left( I_2^{\otimes N} \otimes E_{00} \right) \left( I_2^{\otimes (N-2)} \otimes Q^{(l)} (\xi) \otimes I_2 \right) 
\\
& \qquad \cdots \left( I_2 \otimes Q^{(l)} (\xi) \otimes I_2^{\otimes (N-2)} \right) \left( Q^{(l)} (\xi) \otimes I_2^{\otimes (N-1)} \right)
\\
&
+ \left( I_2^{\otimes (N-1)} \otimes \sigma (\xi) \otimes E_{11} \right) \left( I_2^{\otimes (N-2)} \otimes Q^{(l)} (\xi) \otimes I_2 \right) 
\\
& \qquad \cdots \left( I_2 \otimes Q^{(l)} (\xi) \otimes I_2^{\otimes (N-2)} \right) \left( Q^{(l)} (\xi) \otimes I_2^{\otimes (N-1)} \right)
\\
&= Q^{(g)} _{N} (\xi) \otimes E_{00} + Q^{(g), \sigma} _{N} (\xi) \otimes E_{11}.
\end{align*}
The second equality comes from Eq. \eqref{oto01}. So we have Eq. \eqref{lemoto01}. Noting that $E_{00}^2 = E_{00}, \ E_{11}^2 = E_{11}, \ E_{00} E_{11} = E_{11} E_{00} = O_2$, Eq. \eqref{lemoto01} gives Eq. \eqref{lemoto02}. It follows from ${\rm tr}(A \otimes B) = {\rm tr} (A) {\rm tr} (B)$ and ${\rm tr} (E_{00}) = {\rm tr} (E_{11}) =1$ that Eq. \eqref{lemoto02} implies Eq. \eqref{lemoto03}. Thus, we obtain the desired conclusions.

\hfill$\square$

By Lemma \ref{lemma02}, we have the following result.
\begin{cor}
\begin{align}
& 
Q^{(g), \sigma} _{N+1} (\xi) 
= (- \sin \xi) Q^{(g)} _{N} (\xi) \otimes E_{00} 
+ (\cos \xi) Q^{(g), \sigma} _{N} (\xi) \otimes E_{01} 
\nonumber
\\
& \qquad \qquad + (\cos \xi) Q^{(g)} _{N} (\xi) \otimes E_{10} 
+ (\sin \xi) Q^{(g), \sigma} _{N} (\xi) \otimes E_{11}, 
\label{coroto01}
\\
&
{\rm tr} \left( Q^{(g), \sigma} _{N+1} (\xi) \right) 
= (- \sin \xi) {\rm tr} \left( Q^{(g)} _{N} (\xi) \right) 
+ (\sin \xi) {\rm tr} \left( Q^{(g), \sigma} _{N} (\xi) \right)
\label{coroto02}
\end{align}
for $N, r \in \ZM_{>}$ and $\xi \in [0, 2 \pi)$. Here
\begin{align*}
E_{01} =
\begin{bmatrix}
0 & 1 \\
0 & 0 
\end{bmatrix}
, \quad
E_{10} =
\begin{bmatrix}
0 & 0 \\
1 & 0 
\end{bmatrix}
.
\end{align*}
\end{cor}
{\bf Proof}. By Eq. \eqref{lemoto01}, we observe
\begin{align}
Q^{(g), \sigma} _{N+1} (\xi) 
&= \left( I_2 ^{\otimes N} \otimes \sigma (\xi) \right) Q^{(g)} _{N+1} (\xi) 
\nonumber
\\
&= \left( I_2 ^{\otimes N} \otimes \sigma (\xi) \right) \left( Q^{(g)} _{N} (\xi) \otimes E_{00} + Q^{(g), \sigma} _{N} (\xi) \otimes E_{11} \right)
\nonumber
\\
&= Q^{(g)} _{N} (\xi) \otimes \left( \sigma (\xi) E_{00} \right) 
+  Q^{(g), \sigma} _{N} (\xi) \otimes \left( \sigma (\xi) E_{11} \right). 
\label{kotou01}
\end{align}
On the other hand. 
\begin{align}
\sigma (\xi) E_{00} = (- \sin \xi) E_{00} + (\cos \xi) E_{10}, \quad \sigma (\xi) E_{11} = (\cos \xi) E_{01} + (\sin \xi) E_{11}.
\label{kotou02}
\end{align}
Combining Eq. \eqref{kotou01} with Eq. \eqref{kotou02} gives Eq. \eqref{coroto01}. It follows from ${\rm tr}(A \otimes B) = {\rm tr} (A) {\rm tr} (B)$ and ${\rm tr} (E_{00}) = {\rm tr} (E_{11}) =1, \ {\rm tr} (E_{01}) = {\rm tr} (E_{10}) =0$ that Eq. \eqref{coroto01} implies Eq. \eqref{coroto02}. So the desired conclusions are obtained.

\hfill$\square$

When $N=1$, noting that $Q^{(g)} _{1} (\xi) =I_2$ and $Q^{(g), \sigma} _{1} (\xi) = \sigma (\xi)$, we have
\begin{lemma}
\begin{align}
&
{\rm tr} \left( \left( Q^{(g)} _{1} (\xi) \right)^r \right) = 2,
\nonumber
\\
&
{\rm tr} \left( \left( Q^{(g), \sigma} _{1} (\xi) \right)^r \right) 
= \left\{ 
\begin{array}{ll}
0 & \mbox{$(r \equiv 1 \ ({\rm mod} \ 2))$}, \\
2 & \mbox{$(r \equiv 0 \ ({\rm mod} \ 2))$}
\end{array}
\right.
\label{lemoto002}
\end{align}
for $r \in \ZM_{>}$ and $\xi \in [0, 2 \pi)$.
\end{lemma}

From now on, we first compute $r=1$ case for a general $N \in \ZM_{>}$. To do so, we put
\begin{align*}
x^{(r)} _N = x^{(r)} _N (\xi) = {\rm tr} \left( \left( Q^{(g)} _{N} (\xi) \right)^r \right), \qquad y^{(r)} _N = y^{(r)} _N (\xi) = {\rm tr} \left( \left( Q^{(g), \sigma} _{N} (\xi) \right)^r \right)
\end{align*}
for $r, N \in \ZM_{>}$ and $\xi \in [0, 2 \pi)$. From Eqs. \eqref{lemoto03} ($r=1$) and \eqref{coroto02}, we have
\begin{align}
x^{(1)} _{N+1} 
&= x^{(1)} _{N} + y^{(1)} _{N}, 
\label{kurube01}
\\
y^{(1)} _{N+1} 
&= (- \sin \xi) x^{(1)} _{N} + (\sin \xi) y^{(1)} _{N}.
\nonumber
\end{align}
Thus we get
\begin{align*}
x^{(1)} _{N+2} - \left( 1 + \sin \xi \right) x^{(1)} _{N+1} + 2 (\sin \xi) x^{(1)} _{N} = 0.
\end{align*}
On the other hand, Eq. \eqref{lemoto002} gives $x^{(1)} _{1} = 2$ and $y^{(1)} _{1} = 0$. From Eq. \eqref{kurube01}, we have $x^{(1)} _{2} = 2$. So we get $x^{(1)} _{2} = x^{(1)} _{1} = 2$. Therefore we obtain
\begin{prop}
We assume that $N \in \ZM_{>}$. Let $\lambda_1 = \lambda_1 (\xi)$ and  $\lambda_2 = \lambda_2 (\xi)$ denote solutions of $\lambda^{2} - \left( 1 + \sin \xi \right) \lambda + 2 (\sin \xi)  = 0$. Then the solutions of 
\begin{align*}
x^{(1)} _{N+2} - \left( 1 + \sin \xi \right) x^{(1)} _{N+1} + 2 (\sin \xi) x^{(1)} _{N} = 0
\end{align*}
with $x^{(1)} _{2} = x^{(1)} _{1} = 2$ for $\xi \in [0, 2 \pi)$ are given as follows: 
\par\noindent
{\rm (i)} If $\lambda_1 \not= \lambda_2$, then 
\begin{align*}
{\rm tr} \left( Q^{(g)} _{N} (\xi) \right) 
&= \frac{2}{\lambda_2 - \lambda_1} \Big\{ (\lambda_2 - 1) \lambda_1 ^{N-1} - (\lambda_1 - 1) \lambda_2 ^{N-1} \Big\},
\\
C_1 \left( Q^{(l)} (\xi), \PM_N \right) 
&= \frac{1}{\lambda_2 - \lambda_1} \left\{ (\lambda_2 - 1) \left(\frac{\lambda_1}{2} \right)^{N-1} - (\lambda_1 - 1) \left(\frac{\lambda_2}{2} \right) ^{N-1} \right\}.
\end{align*}
\par\noindent
{\rm (ii)} If $\lambda_1 = \lambda_2$, i.e., $\sin (\xi)=3-2 \sqrt{2}$, then 
\begin{align*}
{\rm tr} \left( Q^{(g)} _{N} (\xi)  \right) 
&= \left\{ \sqrt{2}(N-1) + 2 \right\} \times \left(2 - \sqrt{2} \right)^{N-1}, 
\\
C_1 \left( Q^{(l)} (\xi), \PM_N \right) 
&= \left\{ \frac{\sqrt{2}(N-1) + 2}{2} \right\} \times \left( \frac{2 - \sqrt{2}}{2} \right)^{N-1}.
\end{align*}
\end{prop}
When $\xi=0$ (Rule 90), we see $\lambda_1 = 0$ and $\lambda_2 =1$ (case (i)). Thus we have 
\begin{align}
{\rm tr} \left( Q^{(g)} _{N} (0) \right) = 2, \quad C_1 \left( Q^{(l)} (0), \PM_N \right) = \frac{1}{2^{N-1}}
\label{rule90c1}
\end{align}
for $N \in \ZM_{>}$. When $\xi=\pi/2$, we get $\lambda_1 = 1+i$ and $\lambda_2 =1-i$ (case (i)). So we get
\begin{align*}
&{\rm tr} \left( Q^{(g)} _{N} (\pi/2) \right) = (1+i)^{N-1}+(1-i)^{N-1}, 
\\
&C_1 \left( Q^{(l)} (\pi/2), \PM_N \right) 
= 2^{-(N-1)/2} \ T_{N-1} \left( \frac{\sqrt{2}}{2} \right)
\end{align*}
for $N \in \ZM_{>}$. As a non-trivial case, we take $\xi= \pi/6$, i.e., $\sin (\pi/6)=1/2$ (case (i)). Then we see 
\begin{align*}
\lambda_1 = e^{i \theta} = \frac{3+i \sqrt{7}}{4}, \quad \lambda_2 = e^{- i \theta} = \frac{3-i \sqrt{7}}{4}. 
\end{align*}
Thus we obtain
\begin{align*}
{\rm tr} \left( Q^{(g)} _{N} (\pi/6) \right) 
&= \frac{i \sqrt{7}}{7} \times \left\{ - (1+ i \sqrt{7}) e^{i \theta (N-1)} + (1- i \sqrt{7}) e^{-i \theta (N-1)} \right\}, 
\\
C_1 \left( Q^{(l)} (\pi/6), \PM_N \right) 
&= \left( \frac{1}{2} \right)^{N+1} \left\{ 4 \ T_{N-1} \left( \frac{3}{4} \right) +  \ U_{N-2} \left( \frac{3}{4} \right) \right\}
\end{align*}
for $N \in \ZM_{>}$. Here $\{ U_n (x) \}$ is the Chebychev polynomials of the second kind (see Andrews et al. \cite{Andrews1999}):
\begin{align*}
U_n (\cos \theta) = \frac{\sin ((n+1) \theta)}{\sin \theta} \qquad \left( n \in \ZM_{\ge} \right).
\end{align*}
For example, 
\begin{align*}
U_0 (x) = 1, \quad U_1 (x) = 2x, \quad U_2 (x) = 4x^2-1, \quad U_3 (x) = 8x^3-4x, \ldots.
\end{align*}
Note that we put $U_{-1}(x)=0$.

We next compute $r=2$ case. From Eq. \eqref{lemoto03} ($r=2$), we have 
\begin{align}
x^{(2)} _{N+1} = x^{(2)} _{N} + y^{(2)} _{N}.
\label{oto07}
\end{align}
Define 
\begin{align*}
a_N = Q^{(g)} _{N} (\xi), \qquad b_N = Q^{(g), \sigma} _{N} (\xi). 
\end{align*}
Then Eq. \eqref{coroto01} is rewritten as
\begin{align}
b_{N+1} 
= (- \sin \xi) a_N \otimes E_{00} + (\cos \xi) b_N \otimes E_{01} + (\cos \xi) a_N \otimes E_{10} + (\sin \xi) b_N \otimes E_{11}.
\label{oto08}
\end{align}
So we have
\begin{align*}
\left( b_{N+1} \right)^2 
&= \left\{ (\sin \xi)^2 a_N + (\cos \xi)^2 b_N \right\} a_N \otimes E_{00} + (\sin \xi) (\cos \xi) \left( b_N - a_N \right) b_N \otimes E_{01} 
\\
&+ (\sin \xi)(\cos \xi) \left( b_N - a_N \right) a_N  \otimes E_{10} + \left\{ (\cos \xi)^2 a_N + (\sin \xi)^2 b_N \right\} b_N \otimes E_{11}.
\end{align*}
The trace of the above equation gives
\begin{align}
y^{(2)} _{N+1} = (\sin \xi)^2 x^{(2)} _{N} + 2 (\cos \xi)^2 {\rm tr} \left( a_N b_N \right) + (\sin \xi)^2 y^{(2)} _{N}.
\label{oto09}
\end{align}
Thus we want to compute ${\rm tr} \left( a_N b_N \right)$. Eq. \eqref{lemoto01} is rewritten as
\begin{align}
a_{N+1} = a_N \otimes E_{00} + b_N \otimes E_{11}.
\label{oto09b}
\end{align}
Combining Eq. \eqref{oto08} with Eq. \eqref{oto09b} implies 
\begin{align*}
a_{N+1} b_{N+1} 
&= (- \sin \xi) (a_N)^2 \otimes E_{00} + (\cos \xi) a_N b_N \otimes E_{01} 
\\
&+ (\cos \xi) b_N a_N \otimes E_{10} + (\sin \xi) (b_N)^2 \otimes E_{11}.
\end{align*}
By the trace of the above equation, we get 
\begin{align}
{\rm tr} \left( a_{N} b_{N} \right) = (- \sin \xi) x^{(2)} _{N-1} +  (\sin \xi) y^{(2)} _{N-1}. 
\label{oto10} 
\end{align}
Combining Eq. \eqref{oto09} with Eq. \eqref{oto10} gives
\begin{align}
y^{(2)} _{N+1}= (\sin \xi)^2 x^{(2)} _{N}  + (\sin \xi)^2 y^{(2)} _{N} - 2 (\sin \xi) (\cos \xi)^2 x^{(2)} _{N-1} + 2 (\sin \xi) (\cos \xi)^2 y^{(2)} _{N-1}. 
\label{oto11} 
\end{align}
From Eqs. \eqref{oto07} and \eqref{oto11}, we have
\begin{align}
x^{(2)} _{N+3} - \left\{ 1 + (\sin \xi)^2 \right\} x^{(2)} _{N+2} - 2 (\sin \xi) (\cos \xi)^2 x^{(2)} _{N+1} + 4 (\sin \xi) (\cos \xi)^2 x^{(2)} _{N} = 0
\label{oto12} 
\end{align}
for $n \in \ZM_{>}$ and $\xi \in [0, 2 \pi)$. On the other hand, we compute 
\begin{align*}
x^{(2)} _{1} =2, \quad x^{(2)} _{2} =4, \quad x^{(2)} _{3} = 4 \left\{ 1 + (\sin \xi)^2 \right\}.
\end{align*}
Therefore by Eq. \eqref{oto12}, we obtain
\begin{prop}
We assume that $N \in \ZM_{>}$. Put 
\begin{align*}
x^{(2)} _N = x^{(2)} _N (\xi) = {\rm tr} \left( \left( Q^{(g)} _{N} (\xi) \right)^2 \right)
\end{align*}
for $\xi \in [0, 2 \pi)$. Then we see that $x^{(2)} _N$ is the solution of 
\begin{align}
x^{(2)} _{N+3} - \left\{ 1 + (\sin \xi)^2 \right\} x^{(2)} _{N+2} - 2 (\sin \xi) (\cos \xi)^2 x^{(2)} _{N+1} + 4 (\sin \xi) (\cos \xi)^2 x^{(2)} _{N}= 0
\label{rec02}
\end{align}
with $x^{(2)} _{1} =2, \ x^{(2)} _{2} =4, \  x^{(2)} _{3} = 4 \left\{ 1 + (\sin \xi)^2 \right\}$.
\end{prop}

When $\xi=0$ (Rule 90), we have 
\begin{align*}
x^{(2)} _{N+1} = x^{(2)} _{N} \quad (N=3,4, \ldots).
\end{align*}
Noting that $x^{(2)} _{1} =2, \ x^{(2)} _{2} =4, \  x^{(2)} _{3} = 4$, we see
\begin{align}
&{\rm tr} \left( \left( Q^{(g)} _{N} (0) \right)^2 \right)
=
\left\{ 
\begin{array}{ll}
2 & \mbox{$(N = 1)$}, \\
4 & \mbox{$(N=2,3,\ldots)$},
\end{array}
\right.
\label{rule90c2tr}
\\
&C_2 \left( Q^{(l)} (0), \PM_N \right)
=
\left\{ 
\begin{array}{ll}
\mbox{$\displaystyle \frac{1}{2^{N-1}}$} & \mbox{$(N = 1)$}, 
\\
\\
\mbox{$\displaystyle \frac{1}{2^{N-2}}$} & \mbox{$(N=2,3,\ldots)$}.
\end{array}
\right.
\label{rule90c2}
\end{align}
When $\xi=\pi/2$, we have 
\begin{align*}
x^{(2)} _{N+1} = 2 x^{(2)} _{N} \quad (N=3,4, \ldots).
\end{align*}
Noting that $x^{(2)} _{1} =2, \ x^{(2)} _{2} =4, \  x^{(2)} _{3} = 8$, we get
\begin{align}
{\rm tr} \left( \left( Q^{(g)} _{N} (\pi/2) \right)^2 \right) = 2^N, \quad C_2 \left( Q^{(l)} (\pi/2), \PM_N \right) = 1
\label{pi2c2}
\end{align}
for $N \in \ZM_{>}$. From definitions of $Q^{(l)} (\pi/2)$ and $Q^{(g)} _{N} (\pi/2)$, we easily see that $(k,l)$ element of $Q^{(g)} _{N} (\pi/2)$ is given by 
\begin{align*}
Q^{(g)} _{N} (\pi/2) (k, l) 
=
\left\{ 
\begin{array}{ll}
1 \ \mbox{or} \ -1 & \mbox{$(k=l)$}, \\
0 & \mbox{$(k \not=l)$}.
\end{array}
\right.
\end{align*}
Combining this with Eq. \eqref{pi2c2} implies 
\begin{align*}
\left( Q^{(g)} _{N} (\pi/2) \right)^2 = I_{2^N} \quad (N \in \ZM_{>}).
\end{align*}
Furthermore we have
\begin{align*}
\left( Q^{(g)} _N (\pi/2) \right)^r = I_{2^N}
\end{align*}
for $N \in \ZM_{>}$ and $r \equiv 0 \ ({\rm mod} \ 2)$. In other words, the period of $Q^{(g)} _N (\pi/2)$ is 2. Therefore we obtain
\begin{prop}
We assume that $N \in \ZM_{>}$. Then we have
\par\noindent
{\rm (i)} 
\begin{align}
{\rm tr} \left( \left( Q^{(g)} _{N} (\pi/2) \right)^r \right)
=
\left\{ 
\begin{array}{ll}
\mbox{$\displaystyle 2^{(N+1)/2} \ T_{N-1} \left( \frac{\sqrt{2}}{2} \right)$} & \mbox{$(r \equiv 1 \ ({\rm mod} \ 2))$}, \\
2^N & \mbox{$(r \equiv 0 \ ({\rm mod} \ 2))$}. \\
\end{array}
\right.
\label{pi2crtr}
\end{align}
\par\noindent
{\rm (ii)}
\begin{align}
C_r \left( Q^{(l)} (\pi/2), \PM_N \right)
=
\left\{ 
\begin{array}{ll}
\mbox{$\displaystyle 2^{-(N-1)/2} \ T_{N-1} \left( \frac{\sqrt{2}}{2} \right)$} & \mbox{$(r \equiv 1 \ ({\rm mod} \ 2))$},
\\
\mbox{$\displaystyle 1 $} & \mbox{$(r \equiv 0 \ ({\rm mod} \ 2))$}.
\end{array}
\right.
\label{pi2cr}
\end{align}
\par\noindent
{\rm (iii)}
\begin{align}
\left( Q^{(g)} _N (\pi/2) \right)^r = I_{2^N} \quad (r \equiv 0 \ ({\rm mod} \ 2)).
\label{pi2peri}
\end{align}
\end{prop}
Next we compute the zeta function for this case. To do so, for a constant $c \in \CM$, we observe 
\begin{align*}
&
\frac{c}{1} \ u + \frac{1}{2} \ u^2 + \frac{c}{3} \ u^3 + \frac{1}{4} \ u^4 + \frac{c}{5} \ u^5 + \frac{1}{6} \ u^6 + \cdots
\\
&= c \left( \frac{u}{1} + \frac{u^3}{3} + \frac{u^5}{5} + \cdots \right)
+ \frac{1}{2} \left\{ \frac{(u^2)}{1} + \frac{(u^2)^2}{2} + \frac{(u^2)^3}{3} + \cdots \right\}
\\
&= c ({\rm tanh^{-1}} (u) ) - \frac{1}{2} \log \left( 1 - u^2 \right),
\end{align*}
where ${\rm tanh^{-1}} (u) = {\rm arc tanh} (u)$. By using this, we can compute the zeta function as follows:
\begin{theorem}
We assume that $N \in \ZM_{>}$. Then we have
\begin{align*}
\log \left\{ \overline{\zeta} \left(Q^{(l)}_{QCA,2}(0,\pi/2), \PM_N, u \right)^{-1} \right\} = \frac{1}{2} \log (1-u^2) - 2^{-(N-1)/2} \ T_{N-1} \left( \frac{\sqrt{2}}{2} \right) {\rm tanh^{-1}} (u).
\end{align*}
\label{pi2zeta}
\end{theorem}
By using this theorem, we easily obtain
\begin{cor}
\par\noindent\par\noindent
{\rm (i)} 
\begin{align*}
\lim_{N \to \infty} \log \left\{ \overline{\zeta} \left(Q^{(l)}_{QCA,2}(0,\pi/2), \PM_N, u \right)^{-1} \right\} = \frac{1}{2} \log (1-u) + \frac{1}{2} \log (1+u).
\end{align*}
\par\noindent
{\rm (ii)} 
\begin{align*}
&2^{(N-1)/2} \ T_{N-1} \left( \frac{\sqrt{2}}{2} \right)^{-1} \left[ \frac{1}{2} \log (1-u^2) - \log \left\{ \overline{\zeta} \left(Q^{(l)}_{QCA,2}(0,\pi/2), \PM_N, u \right)^{-1} \right\} \right] 
\\
& \qquad \qquad \qquad \qquad \qquad \qquad \qquad \qquad \qquad \qquad \qquad \qquad \qquad \qquad \qquad 
= {\rm tanh^{-1}} (u).
\end{align*}
\label{pi2zetacor}
\end{cor}
We should remark that part (i) implies that as for eigenvalues of $Q^{(g)}_N$ determined by $Q^{(l)}_{QCA,2}(0,\pi/2)$, the rate of eigenvalues ``$1$" or ``$-1$" converges to $1/2$ as $N \to \infty$.

As a non-trivial case like $r=1$, we consider $\xi= \pi/6$, i.e., $\sin (\pi/6)=1/2$. Then three solutions of the characteristic equation of Eq. \eqref{rec02} are given by
\begin{align*}
\lambda_1 = -1, \quad \lambda_2 = \frac{9+i \sqrt{15}}{8}, \quad \lambda_3 = \frac{9-i \sqrt{15}}{8}.
\end{align*}
Noting that $x^{(2)} _{1} =2, \ x^{(2)} _{2} =4, \  x^{(2)} _{3} = 5$, we obtain
\begin{align*}
&
{\rm tr} \left( Q^{(g)} _{N} (\pi/6) \right)^2 = \left( - \frac{4}{19} \right) \lambda_1^{N-1} + \frac{3}{95} \left( 35 - 11 i \sqrt{15} \right) \lambda_2^{N-1} + \frac{3}{95} \left( 35 + 11 i \sqrt{15} \right) \lambda_3^{N-1}, 
\\
&
C_2 \left( Q^{(l)} (\pi/6), \PM_N \right) = \left( - \frac{2}{19} \right) \left( -2 \right)^{-(N-1)} 
\\
&
\qquad \qquad \qquad \qquad \qquad + \frac{3}{19} \left( \frac{\sqrt{6}}{4} \right)^{N-1} \left\{ 7 \ T_{N-1} \left( \frac{3 \sqrt{6}}{8} \right) + \frac{11 \sqrt{6}}{8} \ U_{N-2} \left( \frac{3 \sqrt{6}}{8} \right) \right\}
\end{align*}
for $N \in \ZM_{>}$. 

Finally we consider a general $r$ case. From now on, we focus on $\xi = 0$ (Rule 90). We prepare two equations. The first one is Eq. \eqref{lemoto03}, i.e.,  
\begin{align}
x^{(r)} _{N+1} = x^{(r)} _{N} + y^{(r)} _{N} \quad (N \in \ZM_{>}). 
\label{isato01}
\end{align}
The second one is Eq. \eqref{oto08}, i.e.,  
\begin{align}
b_{N+1} 
= b_N \otimes E_{01} + a_N \otimes E_{10}.
\label{isato02}
\end{align}
By Eq. \eqref{isato02}, we get
\begin{align}
b_{N+1}^{2s-1} 
&= d_N^{s-1} b_N \otimes E_{01} + c_N^{s-1} a_N \otimes E_{10}, 
\label{isato02a}
\\
b_{N+1}^{2s} 
&= d_N^{s} \otimes E_{00} + c_N^{s} \otimes E_{11}, 
\label{isato02b}
\end{align}
for $s, N \in \ZM_{>},$ where $c_N = a_N b_N, \ d_N = b_N a_N.$ By Eqs. \eqref{isato02a} and \eqref{isato02b}, we have 
\begin{align}
y^{(2s-1)} _{N} 
&= 0 \quad (N=2,3,\ldots),
\label{tsuika01}
\\
y^{(2s)} _{N+1} 
&= 2 \ {\rm tr} \left( c_N^{s} \right) \quad (N \in \ZM_{>}),
\label{tsuika02}
\end{align}
for $s \in \ZM_{>}.$ Combining Eqs. \eqref{isato01} and \eqref{tsuika01} with $x_1^{(2s-1)}=2$ and $y_1^{(2s-1)}=0$ gives 
\begin{align}
x^{(2s-1)} _{N} = 2 \quad (s, N \in \ZM_{>}). 
\label{tsuika03}
\end{align}
Similarly, we obtain
\begin{align}
c_{N+1}^{2s-1} 
= (c_N d_N)^{s-1} c_N \otimes E_{01} + (d_N c_N)^{s-1} d_N \otimes E_{10}, 
\label{tsuika03b}
\end{align}
for $s, N \in \ZM_{>}.$ Thus, from Eqs. \eqref{tsuika02} and \eqref{tsuika03b}, we get
\begin{align}
y^{(2(2s-1))} _{N+2} 
= 2 \ {\rm tr} \left( c_{N+1}^{2s-1} \right) = 0 \quad (s, N \in \ZM_{>}).
\label{tsuika04}
\end{align}
Combining Eqs. \eqref{isato01} and \eqref{tsuika04} with $x_1^{(2(2s-1))} = y_1^{(2(2s-1))} = 2$ and $y_2^{(2(2s-1))} = 0$ implies 
\begin{align}
x^{(2(2s-1))} _{N} = 2^2 \quad (N =2,3, \ldots), \qquad x^{(2(2s-1))} _{1} = 2,
\label{tsuika05}
\end{align}
for $s \in \ZM_{>}.$ In a similar way, we obtain
\begin{align}
x^{(2^2(2s-1))} _{N} 
&= 2^{2^2} \quad (N = 2^2, 2^2+1, \ldots), 
\nonumber
\\
x^{(2^2(2s-1))} _{1} 
&= 2, \quad x^{(2^2(2s-1))} _{2} = 2^2, \quad x^{(2^2(2s-1))} _{3} = 2^3, \quad x^{(2^2(2s-1))} _{4} = 2^4.
\label{tsuika06}
\end{align}
By Eqs. \eqref{tsuika03}, \eqref{tsuika05}, and \eqref{tsuika06}, we have the following result.
\begin{prop}
Put $m \in \{ 1, 2 \}$. For $N=2^{m-1}+1, 2^{m-1}+2, \ldots, 2^m-1, 2^m$, we have
\par\noindent
{\rm (i)} 
\begin{align*}
{\rm tr} \left( \left( Q^{(g)} _{N} (0) \right)^{2^k (2s-1)} \right)
=
\left\{ 
\begin{array}{ll}
2^{2^k} & \mbox{if $k \in \ZM_{\ge}$ with $k < \log_2 N$}, \\
2^N & \mbox{if $k \in \ZM_{\ge}$ with $k \ge \log_2 N$}, \\
\end{array}
\right.
\end{align*}
where $s \in \ZM_{>}.$
\par\noindent
{\rm (ii)}
\begin{align*}
C_{2^k (2s-1)} \left( Q^{(l)} (0), \PM_N \right)
=
\left\{ 
\begin{array}{ll}
\mbox{$\displaystyle \frac{1}{2^{N-2^k}}$} & \mbox{if $k \in \ZM_{\ge}$ with $k < \log_2 N$}, 
\\
\\
\mbox{$\displaystyle 1 $} & \mbox{if $k \in \ZM_{\ge}$ with $k \ge \log_2 N$},
\end{array}
\right.
\end{align*}
where $s \in \ZM_{>}.$
\par\noindent
{\rm (iii)}
\begin{align*}
\left( Q^{(g)} _N (0) \right)^{2^m} = I_{2^N}.
\end{align*}
\end{prop}
Remark that part (iii) implies that the period of $Q^{(g)} _N (0)$ is $2^m$ for $N=2^{m-1}+1, 2^{m-1}+2, \ldots, 2^m-1, 2^m \ (m=1,2)$.

As in the case of $\xi=\pi/2$, we compute the zeta function for $\xi=0$ (Rule 90) in the following way.
\begin{theorem}
Put $m \in \{ 1, 2 \}$. For $N=2^{m-1}+1, 2^{m-1}+2, \ldots, 2^m-1, 2^m$, we have
\begin{align*}
\log \left\{ \overline{\zeta} \left(Q^{(l)}_{QCA,2}(0,0), \PM_N, u \right)^{-1} \right\} 
= \frac{1}{2^m} \log (1-u^{2^m}) - \sum_{k=0}^{m-1}  \frac{1}{2^{N-(2^k-k)}} {\rm tanh^{-1}} (u^{2^k}).
\end{align*}
\label{rule90zeta}
\end{theorem}
\par
Our conjecture is that Theorem \ref{rule90zeta} holds for any $m \in \ZM_{>}$. We should remark that when $N=1$, we see that 
\begin{align*}
\log \left\{ \overline{\zeta} \left(Q^{(l)}_{QCA,2}(0,0), \PM_1, u \right)^{-1} \right\} = \log (1-u).
\end{align*}

\section{Summary \label{sec07}}
The present manuscript defined the two-state discrete time IPS on the one-dimensional path space and introduced the IPS-type zeta function. Section \ref{sec05} treated the zeta function for tensor model, in particular, $Q^{(l)}_{QCA,1}(\xi,\xi) $ case. Furthermore, Section \ref{sec06} dealt with the zeta function for generalized tensor model, especially, $Q^{(l)}_{QCA,2}(0,\xi) $ for $\xi = 0$ (Rule 90), $\xi = \pi/6$, and $\xi = \pi/2$ cases. This paper is the first step for the study of ``IPS/Zeta Correspondence". One of the interesting future problems might be to extend the one-dimensional lattice to a suitable class of general graphs.


\end{document}